\documentclass[a4paper,11pt]{article}
\usepackage{pos}
\usepackage[normalem]{ulem}
\usepackage{comment}
\usepackage{slashed,ulem}

\newcommand{\be}{\begin{equation}}
\newcommand{\ee}{\end{equation}}
\newcommand{\bea}{\begin{eqnarray}}
\newcommand{\eea}{\end{eqnarray}}

\newcommand{\MSbar}{{\overline{\rm MS}}}

\newcommand{\qslash}{{\not{\hspace{-0.05cm}q}}}

\newcommand{\la}{\lambda}

\newcommand{\al}{\alpha}

\def\openone{\leavevmode\hbox{\small1\kern-6.8pt\normalsize1}}

\title{First Overlap Formulation and Perturbative Results for Supersymmetric QCD}

\author*[a]{Marios Costa}
\author[b]{Elias Antoniou}
\author[b]{Herodotos Herodotou}
\author[c]{Eleni Ioannou}
\author[c]{Constantinos Kartoudes}
\author[c]{Gregoris Spanoudes}
\author[c]{Haralambos Panagopoulos}

\affiliation[a]{Department of Mechanical Engineering and Material Science and Engineering, Cyprus University of Technology, Limassol, CY-3036, Cyprus}
\affiliation[b]{Rinnoco Ltd, Limassol, CY-3047, Cyprus}
\affiliation[c]{Department of Physics, University of Cyprus, Nicosia, CY-1678, Cyprus}

\emailAdd{marios.costa@cut.ac.cy}
\emailAdd{elias.a@rinnoco.com}
\emailAdd{herodotos.h@rinnoco.com}
\emailAdd{ioannou.c.eleni@ucy.ac.cy}
\emailAdd{kartoudes.constantinos@ucy.ac.cy}
\emailAdd{spanoudes.gregoris@ucy.ac.cy}
\emailAdd{panagopoulos.haris@ucy.ac.cy}

\abstract{We present a lattice formulation of 
 $\mathcal{N}=1$ supersymmetric QCD (SQCD) based on overlap fermions, providing a discretization consistent with the Ginsparg–Wilson relation. Particular emphasis is placed on the construction of Yukawa terms that preserve a lattice-modified chiral symmetry involving overlap quarks and gluinos. To realize this modified chiral symmetry in the Yukawa sector, we follow Lüscher’s strategy and introduce auxiliary Dirac and Majorana fermionic fields for the quark and gluino fields, respectively. After functional integration over these auxiliary fields, ultralocal interaction terms arise as additional contributions to the lattice action. Within this framework, and using the Wilson gauge action for gluons, we also carry out one-loop perturbative calculations. In particular, we compute the self-energies of the quark, gluon, gluino, and squark fields in the presence of quark and squark masses, enabling an unambiguous determination of the corresponding field and mass counterterms on the lattice. This work provides a robust foundation for future perturbative determinations of the fine-tunings of the gauge, Yukawa, and quartic couplings, and already shows, as expected, a reduction in the number of required counterterms relative to Wilson-type fermions.}

\FullConference{The 43rd International Symposium on Lattice Field Theory (LATTICE2026)\\
July 26 to August 1, 2026,\\
University of Maryland, College Park, USA\\}

\begin{document}
\maketitle

\vspace{-0.75cm}
\section{Introduction and motivation}
\label{sec:introduction}
\vspace{-0.25cm}
Lattice regularization provides a well-defined framework for the nonperturbative study of gauge theories. Extending lattice techniques to supersymmetric theories is challenging, as it would enable first-principles investigations of confinement, chiral symmetry breaking, supersymmetry (SUSY) restoration, and the spectrum of bound states in theories beyond the Standard Model. In particular, four-dimensional $\mathcal{N}=1$ supersymmetric QCD (SQCD) constitutes one of the simplest nontrivial supersymmetric extensions of QCD, containing gluons together with their fermionic superpartners (gluinos), as well as massive matter multiplets consisting of quarks and squarks.

A central obstacle in lattice formulations of supersymmetric theories is that supersymmetry is generically broken by the lattice regularization at nonzero lattice spacing~\cite{Curci:1986sm}. Consequently, recovering supersymmetry in the continuum limit requires fine tuning of the bare lattice action. The difficulty is particularly acute in SQCD, where the enlarged field content allows a large number of relevant and marginal operators~\cite{Giedt:2008xm,Costa:2024fvy}. In Wilson-type formulations, the explicit breaking of chiral symmetry leads to additive mass renormalizations and the generation of additional Yukawa and quartic scalar interactions, substantially increasing the number of required counterterms~\cite{Giedt:2008xm,Costa:2024fvy}.

Overlap fermions provide an attractive alternative. By satisfying the Ginsparg--Wilson relation, they realize an exact lattice-modified chiral symmetry~\cite{Luscher:1998pqa}, protecting fermion masses from additive renormalization and suppressing several symmetry-breaking counterterms. This reduction in the number of required fine tunings makes overlap fermions particularly well suited for lattice studies of SQCD.

In these proceedings, we summarize recent progress in the formulation of $\mathcal{N}=1$ SQCD with overlap fermions. Following Lüscher's strategy, auxiliary Dirac and Majorana fermion fields are introduced in order to construct Yukawa interactions compatible with the lattice-modified chiral symmetry. We also present the first one-loop perturbative results within this overlap formulation, including the self-energies of quarks ($\psi$), squarks ($A_\pm$), gluons ($u_\mu$), gluinos ($\lambda$), and ghosts ($c$) as well as their corresponding field and mass renormalization factors. These results provide a first step toward the perturbative determination of the fine tunings required for future nonperturbative simulations of supersymmetric QCD on the lattice.

\vspace{-0.2cm}
\section{Overview of the lattice formulation}
\label{sec:setup}
Our starting point is the continuum action of $\mathcal{N}=1$ SQCD with gauge group $SU(N_c)$ and $N_f$ matter multiplets, containing gluons, gluinos, quarks and squarks. The continuum theory includes gauge, fermion and scalar kinetic terms, Yukawa interactions, quartic squark couplings and a common quark--squark mass term required by supersymmetry~\cite{Wess:1992cp, Martin:1997ns, Costa:2017rht}.

To formulate the theory on the lattice while preserving chiral symmetry, we employ overlap fermions for both quark and gluino fields. The quark action is obtained by replacing the continuum Dirac operator with the overlap operator $\mathcal{D}_{\rm ov}$ in the fundamental representation, while retaining the massive overlap form for nonzero quark masses ~\cite{Niedermayer:1998bi, Alexandrou:1999wr, Alexandrou:2000kj}. For the gluino sector, the overlap construction is implemented in the adjoint representation through the operator $\mathcal{D}^{\rm adj}_{\rm ov}$, taking into account the Majorana nature of the gluino field. Both overlap operators satisfy the Ginsparg--Wilson relation and therefore realize an exact lattice-modified chiral symmetry \cite{Clancy:2023ino}. Components of the SQCD action involving only gluons and squarks admit a more standard discretization. For the gluon sector we employ the Wilson plaquette gauge action, while the squark kinetic terms are constructed using nearest-neighbor covariant lattice derivatives. The squark mass and quartic scalar interactions being ultralocal, their discretizations are also standard.  The definitions of $\mathcal{D}_{\rm ov}$, $\mathcal{D}^{\rm adj}_{\rm ov}$, as well as those of all other components entering the lattice actions, can be found in Refs.~\cite{Costa:2025vnq, Costa:2026}. We have also introduced the Wilson parameters ($r^\psi$, $r^\lambda$) and the overlap parameters ($\rho^\psi$, $\rho^\lambda$) that enter the calculation, where the superscripts $\psi$ and $\lambda$ refer to the quark and gluino fields, respectively.

The main difficulty arises in the Yukawa sector. Although the remaining terms of the overlap fermion action are invariant under the lattice chiral transformations associated with the Ginsparg--Wilson relation, supersymmetric Yukawa interactions are not invariant under the lattice transformations. To overcome this problem, we follow the construction proposed by Lüscher~\cite{Luscher:1998pqa} and introduce two non-dynamical auxiliary fermion fields, which vanish in the continuum limit: a Dirac quark field $\chi_\psi$ in the fundamental representation and a Majorana gluino field $\chi_\lambda$ in the adjoint representation. 
The Yukawa interactions are then modified  via substitutions: 
$\psi \to (\psi+\chi_\psi)$ and $\lambda \to(\lambda+\chi_\lambda)$\,. Furthermore, the auxiliary
fields acquire an additional quadratic action of the form:
\begin{equation}
\mathcal{S}_{\rm aux}
=
a^4 \sum_x
\left[
-\frac{\rho^\lambda}{a}\,
\bar\chi_\lambda^\alpha \chi_\lambda^\alpha
-\frac{2\rho^\psi}{a}\,
\bar\chi_\psi \chi_\psi
\right].
\end{equation}

The auxiliary fields are assigned modified chiral transformation properties such
that the combinations $(\psi+\chi_\psi)$ and $(\lambda+\chi_\lambda)$ transform
as ordinary continuum chiral fields. In particular,
\begin{equation}
\delta\psi =
\gamma_5\left(1-\frac{a\mathcal{D}_{\rm ov}}{2\,\rho^\psi}\right)\psi
+\gamma_5\chi_\psi,
\quad
\delta\chi_\psi =
\gamma_5\frac{a\mathcal{D}_{\rm ov}}{2\,\rho^\psi}\psi,
\quad
\delta(\psi+\chi_\psi) =
\gamma_5(\psi+\chi_\psi).
\end{equation}
Analogous transformations apply to the gluino and its corresponding auxiliary
field. Consequently, the Yukawa action is exactly invariant under the modified
lattice chiral symmetry, and so is the total action.

Since the auxiliary fields appear only quadratically and linearly in the action, they can be integrated out exactly. Performing the functional integration generates additional contributions to the lattice action, including effective ultralocal interaction terms and a squark-dependent Pfaffian contribution. The resulting formulation provides a chirally symmetric lattice realization of $\mathcal{N}=1$ SQCD and serves as the basis for the perturbative calculations presented in the following sections. The total action of SQCD in the overlap lattice formulation reads:
\begin{equation}
{\cal S}^{\rm Total}_{\rm SQCD} = {\cal S}^{L}_{\rm SQCD} + \mathcal{S}_{\lambda\lambda} + \mathcal{S}_{\text{Pf}} \,+ \mathcal{S}_{\Psi\Psi} \, ,
\end{equation}
where the terms ${\cal S}^{L}_{\rm SQCD}$, $\mathcal{S}_{\lambda\lambda}$, $\mathcal{S}_{\text{Pf}}$ and $\mathcal{S}_{\Psi\Psi}$ are shown in Ref.~\cite{Costa:2025vnq}. 
These three additional contributions are ultra-local, arise at higher orders in
the coupling constant $g$, and vanish in the continuum limit, $a\to0$. They
also contain non-polynomial terms in the squark fields.

\vspace{-0.2cm}
\subsection{Perturbative setup}
In order to avoid divergences from the integration over gauge orbits, we add two standard terms in the action: the gauge-fixing term, $S_{GF}^{L}$, and the compensating ghost field term, $S_{Ghost}^{L}$; these terms are the same as in the non-supersymmetric case. Furthermore, an additional term, the ``measure'' term $S_{M}^{L}$, must be added to the lattice action in order to account for the Jacobian in the change of integration variables: $U_\mu \to u_\mu$\,. 

Having constructed the overlap formulation of SQCD in Ref.~\cite{Costa:2025vnq}, our next step in Ref.~\cite{Costa:2026} is to develop the corresponding perturbative framework for calculating the one-loop lattice corrections to the propagators of the theory. To this end, the lattice action is expanded in powers of the bare lattice coupling $g$, including terms up to order $g^2$. The perturbative expansion of overlap quarks in QCD and overlap gluinos in supersymmetric Yang--Mills theory has been extensively studied in the literature, see e.g., Refs.~ \cite{Constantinou:2007rm, Constantinou:2007gv}. Therefore, most of the vertices originating from the overlap fermion actions and from the Wilson gauge action are already known. Hence, we only need to focus on the perturbative expansion of the additional contributions generated after integrating out the auxiliary fields introduced to preserve chiral symmetry in the Yukawa sector ($\mathcal{S}_{\lambda\lambda}$, $\mathcal{S}_{\text{Pf}}$ and $\mathcal{S}_{\Psi\Psi}$).

The first one, $S_{\lambda\lambda}$, is already expressed as a polynomial in $g$ and can be used directly in perturbation theory:
\be
\mathcal{S}_{\lambda\lambda} =
\frac{a\,g^2}{2} a^4\sum_x
\bar\lambda^\alpha(x)
\Big[
\left(A_+^\dagger(x)\{T^\alpha,T^\beta\}A_-^\dagger(x)\right)P_+
+
\left(A_-(x)\{T^\alpha,T^\beta\}A_+(x)\right)P_-
\Big]
\lambda^\beta(x).
\ee
The second contribution, $S_{\rm Pf}$, is expanded in powers of the coupling, generating additional scalar interaction vertices:
\begin{align}
\mathcal{S}_{\text{Pf}} &=  \frac{a\,g^2}{2\mu} \Bigl(\frac{N_c^2-1}{N_c}\Bigr) \Bigl((A_+^{\dagger f} A_-^{\dagger f})+(A_-^f A_+^f)\Bigr)  + \mathcal{O}(g^4).
\label{SPfperturbative}
\end{align}
Finally, the perturbative expansion of $S_{\Psi\Psi}$ is shown below: 
\bea
    \mathcal{S}_{\Psi \Psi} &=& \frac{1}{4}\,a^4 \sum_x \bigg\{ \frac{2\,a\,g^2}{\rho^\la} \Big[2\left( (\bar \psi T^\al A_+) \,P_-\, (A_- T^\al \psi) + (\bar \psi T^\al A_-^\dagger)\, P_+\, (A_+^\dagger T^\alpha \psi) \right) \nonumber \\ &&
    - \big((\bar \psi T^\al A_+) \, P_- C^\dagger\,  (A_+ {T^\al}^T {\bar \psi}^T) 
   + (\bar \psi T^\al A_-^\dagger)\, P_+ C^\dagger\, (A_-^\dagger {T^\al}^T {\bar \psi}^T) \big) 
   \nonumber \\ &&
   + \left((\psi^T {T^\al}^T A_+^\dagger)\, C P_+ \, (A_+^\dagger T^\al \psi) + (\psi^T {T^\al}^T A_-) \, C P_-\, (A_- T^\al \psi) \right) \Big]\bigg\} + \mathcal{O}(g^3).
\eea
A noteworthy feature of ${\mathcal S}_{\Psi\Psi}$ is that it introduces coupling among squarks and a pair of
quarks, or a pair of anti-quarks.
\vspace{-0.5cm}
\section{Self energies of all fields and Results on Renormalization factors}

As a first application of the perturbative framework, we compute the one-loop self-energies of all elementary fields in the theory. The corresponding one-particle-irreducible two-point Green's functions are evaluated using lattice regularization, including all one-loop diagrams generated by the overlap formulation. From these results we extract the field and mass renormalization factors by matching the lattice Green's functions to the corresponding renormalized expressions in the $\MSbar$ scheme. The renormalized ($R$) fields and mass parameters are related to their bare ($B$) counterparts through the corresponding renormalization factors:
\bea
u_{\mu}^R &=& \sqrt{Z_u}\,u^B_{\mu},\quad
\la^R = \sqrt{Z_\la}\,\la^B,\quad
\psi^R = \sqrt{Z_\psi}\,\psi^B,\quad
A^R = \sqrt{Z_A}\,A^B, \nonumber\\
m_\psi^R &=& Z_{m_\psi}\,m_\psi^B,\quad
{{m_A^R}}^\dagger m_A^R =
Z_{m_A}^\dagger {{m_A^B}}^\dagger m_A^B Z_{m_A},\quad
c^R = \sqrt{Z_c}\,c^B,
\eea
where the squark field components are arranged into a doublet: $ A \equiv \begin{pmatrix} A_+ \\ A_-^\dagger \end{pmatrix}$ and, correspondingly, $Z_{m_A}$ is a
$2\times 2$ matrix.

\begin{figure}[ht!]
\centering
\includegraphics[scale=0.17]{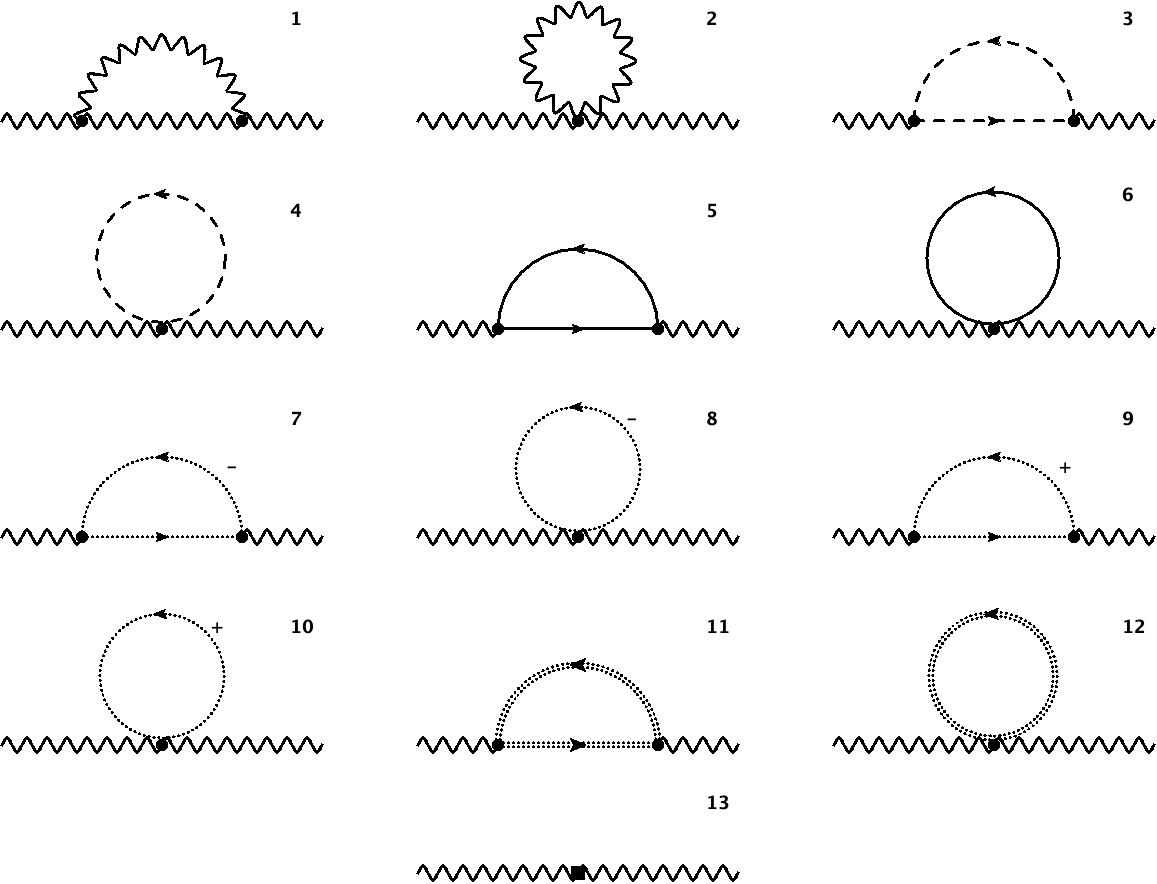}
\caption{One-loop diagrams for the gluon two-point Green's function. A wavy line denotes gluons, a dashed line denotes gluinos, a dotted line denotes squarks, a solid line denotes quarks, a double dotted line denotes the ghost field. The square vertex (bottom diagram) denotes contributions arising from the integration measure.}
\label{fig2ptGG}
\end{figure}

We now turn to the calculation of the self-energies. We begin by presenting the one-loop gluon inverse propagator, obtained by summing the diagrams shown in Fig.~\ref{fig2ptGG} that contribute to the inverse ($\langle u_\mu u_\nu\rangle_{\rm inv}$) of the two-point Green's function $\langle u_\mu  u_\nu \rangle$.

\vspace{-0,25cm}
For $r^\lambda = r^\psi= 1$, the one-loop gluon inverse propagator takes the form:
\be
\big\langle \tilde u^\alpha_\mu (q)\,\tilde u^\beta_\nu(q')  \big\rangle^{LR}_{\rm inv} =
(2\pi)^4 \delta(q+q') \ \frac{\delta^{\alpha \beta}}{2} \Bigg[\frac{1}{\alpha} q_{\mu} q_{\nu} + \left(q^2 \delta_{\mu \nu} - q_{\mu} q_{\nu}\right)\left(1 - \Sigma^{\rm latt}_u (q,m_A,m_\psi)\right) \Bigg],
\ee
where the gluon lattice self-energy is:
\vspace{-0,25cm}
\begin{align}
\hspace{-0.5cm}\Sigma^{\rm latt}_u(q,m_A,m_\psi) = \frac{g^2}{16\pi^2}\Bigg[N_f \Bigg(&c_u^{N_f} - \frac{4}{3q^3}(4m_A^2 + q^2)^{\frac{3}{2}} \, {\rm arctanh}\left(\frac{q}{\sqrt{4m_A^2+q^2}}\right) \nonumber \\ 
& + \frac{8}{3q^3}\sqrt{4m_\psi^2 + q^2} \, (2m_\psi^2-q^2) \, {\rm arctanh} \left(\frac{q}{\sqrt{4m_\psi^2+q^2}}\right) \nonumber \\ 
& + \frac{16}{3q^2}(m_A^2 - m_\psi^2) - \frac{2}{3}\log(a^2m_A^2) - \frac{4}{3}\log(a^2m_\psi^2)\Bigg) \nonumber \\ 
&  + \frac{39.4784}{N_c}  + N_c\left(c_u^{N_c} + 1.7726\alpha - \frac{1}{2}\alpha^2 + (3-\alpha)\log(a^2q^2)\right)\Bigg],
\end{align}
and the finite lattice coefficients $c_u^{N_f}$ and $c_u^{N_c}$ depend on the parameters $\rho^\psi$ and $\rho^\lambda$, respectively. Here, \(m_\psi\) and \(m_A\) denote the bare quark and squark masses, respectively, which generally differ, although their renormalized counterparts are equal. We have obtained results for overlap parameter values ranging from \(0.1\) to \(1.9\), in steps of \(0.1\). The inverse gluon propagator is transverse, with the longitudinal terms vanishing when combinatorial factors are introduced. Furthermore, no gluon critical mass is generated as terms proportional to $1/a^2$ also cancel out.
The gluon field renormalization factor, $Z_u$, is determined by:
\be
\langle \Tilde{u}^\alpha_\mu(q) \, \Tilde{u}^\beta_\nu(q') \rangle^{\MSbar}_{\rm inv} = \left[Z_u^{LR,\MSbar} \right]^{-1} \langle \Tilde{u}^\alpha_\mu(q) \, \Tilde{u}^\beta_\nu(q') \rangle^{LR}_{\rm inv}.
\ee
Setting \(r^{\psi}=\rho^{\psi}=r^{\lambda}=\rho^{\lambda}=1\), we obtain the following results at one-loop order:
\be
Z_u^{LR,\MSbar} = 1 - \frac{g^2}{16\pi^2} \Big[N_c \left(43.7766 - 2.7726\alpha - (3-\alpha)\log(a^2 \bar \mu^2) \right) 
+ \frac{39.4784}{N_c} - N_f \left(4.7247 + 2 \log(a^2 \bar \mu^2) \right) \Big].
\ee

\begin{figure}
\centering
\includegraphics[width=0.375\textwidth]{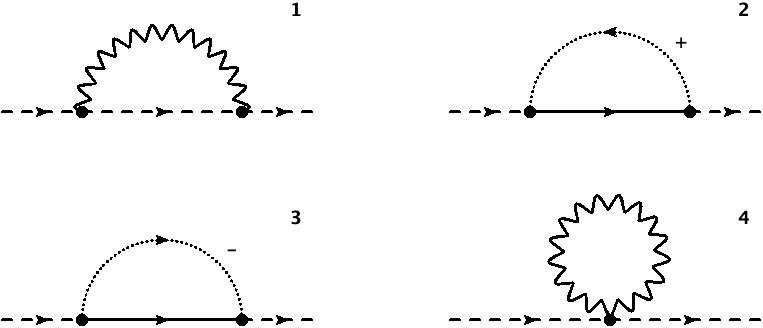}
\caption{One-loop diagrams for the gluino two-point Green's function. Notation is identical to that of Figure \ref{fig2ptGG}.}
\label{fig2ptgg}
\end{figure}

We next consider the gluino inverse propagator, which is obtained at one-loop order by summing the diagrams shown in Fig.~\ref{fig2ptgg} that contribute to the two-point Green's function $\langle \lambda \bar{\lambda} \rangle$. The one-loop inverse gluino propagator takes the form:
\be
\big\langle \tilde \lambda^\alpha(q)
\tilde{\bar{\lambda}}^\beta(q') \big\rangle^{LR}_{\rm inv}
=
(2\pi)^4 \delta(q-q')\,
i\qslash\,
\frac{\delta^{\alpha\beta}}{4}
\left[
1-\Sigma_\lambda^{\rm latt}(q,m_A,m_\psi)
\right].
\ee
where the gluino lattice self-energy is:
\begin{align}
\Sigma^{\mathrm{latt}}_\la(q,m_A,m_\psi)
&= \frac{g^2}{16\pi^2} \Bigg[N_f \Bigg(c^{N_f}_{\lambda} + \log\!\left(a^2 q^2\right) - \frac{1}{q^2} \left(m_A^2 - m_\psi^2 \right) + N_c \left( \alpha \log(a^2 q^2) + c^{N_c}_{\lambda}  - 4.79201 \, \alpha \right) 
\nonumber\\[6pt]
&+ \frac{1}{2 (q^2)^2} \left((m_A^2 - m_\psi^2)^2 + 2 m_A^2 q^2 \right)
\log\!\left(\frac{m_A^2}{m_\psi^2}\right)
- \log\!\left(\frac{q^2}{m_A m_\psi}\right)
\nonumber\\[6pt]
&\quad + \frac{m_A^2 - m_\psi^2 + q^2}{2 (q^2)^2}
\sqrt{(m_A^2 - m_\psi^2)^2 + 2(m_A^2 + m_\psi^2)q^2 + (q^2)^2} \, \times
\nonumber\\
&\quad
\log\!\left(
\frac{m_A^2 + m_\psi^2 + q^2 + \sqrt{(m_A^2 - m_\psi^2)^2 + 2(m_A^2 + m_\psi^2)q^2 + (q^2)^2}}
{m_A^2 + m_\psi^2 + q^2 - \sqrt{(m_A^2 - m_\psi^2)^2 + 2(m_A^2 + m_\psi^2)q^2 + (q^2)^2}}
\right) \Bigg)\Bigg],
\label{sigmala}
\end{align}
$c_\lambda^{N_f}$ and $c_\lambda^{N_c}$ are numerical coefficients that depend on the parameters ($r^\psi$,
$\rho^\psi$) and ($r^\lambda$, $\rho^\lambda$), respectively. For the specific cases $r^\psi = \rho^\psi = 1$ and $r^\lambda = \rho^\lambda = 1$, the coefficients take the following values:
\be
c^{N_f}_\lambda = -2.07534, \qquad c^{N_c}_\lambda = -32.8386.
\ee
As follows from Eq.~\eqref{sigmala}, no gluino critical mass is generated, as expected. Furthermore, although individual diagrams contain terms proportional to $\gamma_5\qslash$, these contributions cancel upon summation. From this result, we extract the gluino field renormalization factor, $Z_\la$, using the following condition:
\be
\big\langle
\tilde \lambda^\alpha(q)
\tilde{\bar{\lambda}}^\beta(q')
\big\rangle^{\MSbar}_{\rm inv}
=
\left [Z_\la^{LR,\MSbar}\right]^{-1} \big\langle
\tilde \lambda^\alpha(q)
\tilde{\bar{\lambda}}^\beta(q')
\big\rangle^{LR}_{\rm inv}.
\ee
At one-loop order, setting $r^{\psi}=\rho^{\psi}=r^{\lambda}=\rho^{\lambda}=1$, we obtain:
\be
Z_\lambda^{LR,\MSbar}
=
1
-
\frac{g^2 N_c}{16\pi^2}
\left[
\alpha
\left(
\log(a^2\bar{\mu}^2)-3.79201
\right)
+
c_\lambda^{N_c}
\right]
-
\frac{g^2 N_f}{16\pi^2}
\left[
2+\log(a^2\bar{\mu}^2)
+
c_\lambda^{N_f}
\right].
\ee

\begin{figure}
\centering
\includegraphics[width=0.375\textwidth]{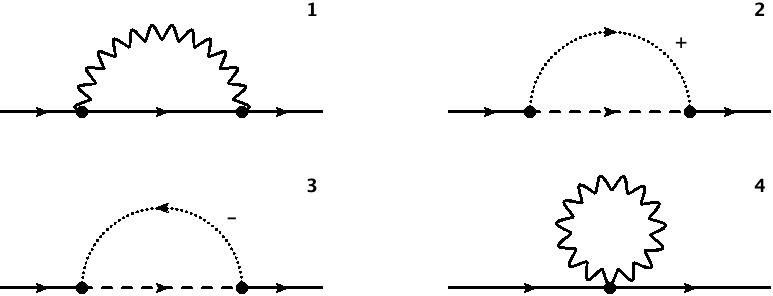}
\caption{One-loop diagrams for the quark two-point Green's function. 
}
\label{fig2ptpp}
\end{figure}

We now turn to the quark propagator, for which the corresponding Feynman diagrams are shown in Fig.~\ref{fig2ptpp}, contributing to the two-point Green's function $\langle \psi \bar{\psi} \rangle$.

The one-loop quark inverse propagator takes the form:
\be
\big\langle \tilde \psi(q) \tilde{\bar{\psi}}(q') \big\rangle^{LR}_{\rm inv}
= (2\pi)^4 \delta(q-q') \left[(i \qslash-m_\psi) -\Sigma_\psi^{\rm latt}(q,m_A,m_\psi)\right],
\ee
where $\Sigma_{\psi}^{\rm latt}$ is as follows for $r^\psi=r^\la=1$:
\bea
\hspace{-1.5cm}\Sigma_\psi^{\rm latt}(q,m_A,m_\psi) &=& \frac{g^2 C_F}{16 \pi^2} \Bigg[ i \qslash \Bigg(
c^{\qslash}_\psi
+ 4.7920\,\alpha
+ \frac{m_A^2}{q^2} 
- \alpha \frac{m_\psi^2}{q^2} - \left(\frac{m_A^4}{{(q^2)}^2}
+ 2 \frac{m_A^2}{q^2}\right)\log\!\left(1+\frac{q^2}{m_A^2} \right) + \log\!\left(\frac{1}{a^2(m_A^2 + q^2)} \right) \nonumber\\
&&\hspace{2cm} + \alpha \log\!\left(\frac{1}{a^2(m_{\psi}^2 + q^2)} \right)
- \alpha \frac{m_\psi^4}{{(q^2)}^2}\log\!\left(1+\frac{q^2}{m_\psi^2} \right)
\Bigg)\nonumber\\
&&\hspace{1cm} - m_\psi \Bigg(
c^{m_\psi}_\psi
+ 5.7920\,\alpha + (3 + \alpha)\log\!\left(\frac{1}{a^2 (m_\psi^2 + q^2)}\right) 
- (3+\alpha) \frac{m_\psi^2}{q^2}\log\!\left(1+\frac{q^2}{m_\psi^2}\right)\Bigg) \Bigg],
\label{InvQuarkprop}
\eea
where $C_F=(N_c^2-1)/(2\,N_c)$ is the quadratic Casimir operator in the fundamental representation. The coefficients $c^{m_\psi}_\psi$ and $c^{\qslash}_\psi$ from Eq.~\eqref{InvQuarkprop} depend on the values of $\rho^\psi$ and $\rho^\la$. What is noteworthy is that while $c_\psi^{m_\psi}$ only depends on $\rho_\psi$, since contributions proportional to $m_\psi$ only arise from diagram 1 of Fig.~\ref{fig2ptpp}, $c_\psi^{\qslash}$ on the other hand depends on both $\rho^\psi$ and $\rho^\la$ since all four diagrams of Fig.~\ref{fig2ptpp} can contribute to terms proportional to $\qslash$.

A suitable condition for the renormalization of the quark propagator is the following:
\be
\big\langle \tilde \psi(q) \tilde{\bar{\psi}}(q') \big\rangle_{\rm{inv}}^{\MSbar} = \left [Z_{\psi}^{LR, \MSbar} \right]^{-1} \big\langle \tilde \psi(q) \tilde{\bar{\psi}}(q') \big\rangle_{\rm{inv}}^{LR}.
\ee
In particular, from the above equation, two conditions can be derived by considering separately the terms proportional to $\slashed{q}$ and $m_\psi$:
\be
    \left[Z_{\psi}^{LR,\MSbar} \right]^{-1} i \slashed q - \left(\Sigma_{\psi}^{\rm latt} - \Sigma_{\psi}^B \right) \Big|_{\propto \ i \slashed q}  = i \slashed q, \quad
    \left[Z_{\psi}^{LR,\MSbar} \right]^{-1} \left[Z_{m_{\psi}}^{LR,\MSbar} \right]^{-1} m_{\psi}^R + \left(\Sigma_{\psi}^{\rm latt} - \Sigma_{\psi}^B \right) \Big|_{\propto \ m_{\psi}}  = m_{\psi}^R. 
\ee

This allows us to extract the renormalization factors for the quark field and the quark mass. For
$\rho^\lambda = \rho^\psi = r^\lambda = r^\psi = 1$, the quark field renormalization factor, $Z_\psi$, is given by:
\begin{equation}
    Z_{\psi}^{LR,\MSbar}
    = 1 - \frac{g^2 C_F}{16 \pi^2}
    \left[
        32.9140 + 3.7920\,\alpha
        - (1+\alpha)\log(a^2 \bar{\mu}^2)
    \right],
\end{equation}
while, as expected, the renormalization of the quark mass, $Z_{m_\psi}$, is gauge independent:
\begin{equation}
    Z_{m_{\psi}}^{LR,\MSbar}
    = 1 - \frac{g^2 C_F}{16 \pi^2}
    \left[
        -32.4440 - \log(a^2 \bar{\mu}^2)
    \right].
\end{equation}

\begin{figure}[ht!]
\centering
\begin{minipage}{0.48\textwidth}
\centering
\includegraphics[scale=0.15]{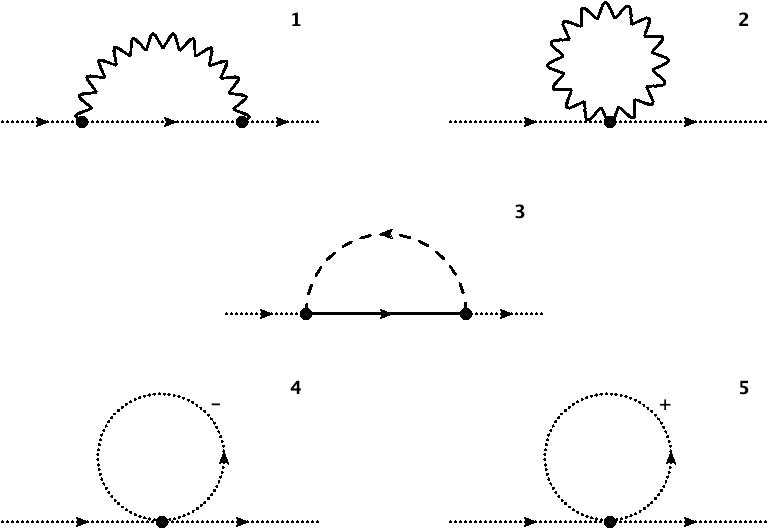}
\caption{One-loop Feynman diagrams contributing to the diagonal two-point
Green's functions \(\langle A_+ A^\dagger_+ \rangle\) and
\(\langle A^\dagger_- A_- \rangle\).}
\label{fig2ptSS}
\end{minipage}
\hfill
\begin{minipage}{0.48\textwidth}
\centering
\includegraphics[scale=0.15]{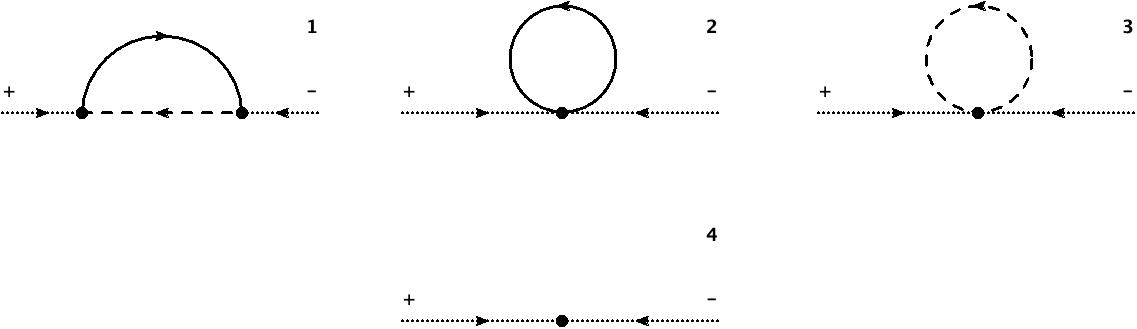}
\caption{One-loop Feynman diagrams contributing to the nondiagonal two-point
Green's functions \(\langle A_-^\dagger A_+^\dagger \rangle\) and
\(\langle A_+ A_- \rangle\).}
\label{fig2ptS+S-}
\end{minipage}
\end{figure}

For the squark propagator, it is convenient to arrange its components into a doublet. The diagonal part of the Green's function
\(\big\langle \tilde A(q),\tilde A^{\dagger}(q') \big\rangle^{LR}_{\rm inv}\)
is obtained from the sum of the diagrams shown in Fig.~\ref{fig2ptSS},
whereas its nondiagonal part is obtained from the diagrams shown in
Fig.~\ref{fig2ptS+S-}.

\vspace{-0.5cm}
The inverse squark propagator takes the form:
\begin{align}
\big\langle \tilde A(q) {\tilde A}^{\dagger}(q') \big\rangle^{LR}_{\rm inv}
= (2\pi)^4 \delta(q-q') \Bigg[& (q^2 + m_A^2)
- \Sigma_A^{\rm latt}(q,m_A,m_\psi)
 \Bigg] \begin{pmatrix} 1 & 0\\ 0 & 1 \end{pmatrix},
\end{align}
\vspace{-0,25cm}
where the off-diagonal contribution vanishes and,
\vspace{-0,05cm}
\begin{align}
\Sigma_A^{\rm latt}(q, m_A, m_\psi)
&=
\frac{g^2 C_F}{16\pi^2}
\Bigg[
\frac{c_{-2}}{a^2}
+ \frac{c_{-1}m_\psi}{a}
+ c_{m_\psi^2}m_\psi^2 
+ m_A^2 \Big(
1.7920 - 3.7920\,\alpha
\Big)
+ q^2 \Big(
c_{q^2} - 3.7920\,\alpha
\Big) \nonumber\\
&\quad
-
\Big(
(1-\alpha)m_A^2 + (3-\alpha)q^2
\Big)
\log\!\big(a^2 m_A^2\big)
+
\Big(
4m_\psi^2 + 2q^2
\Big)
\log\!\big(a^2 m_\psi^2\big) \nonumber\\
&\quad
+ (3-\alpha)
\Big(
\frac{m_A^4}{q^2} - q^2
\Big)
\log\!\left(1+\frac{q^2}{m_A^2}\right) 
+
\Big(
\frac{2m_\psi^4}{q^2} + 4m_\psi^2 + 2q^2
\Big)
\log\!\left(1+\frac{q^2}{m_\psi^2}\right)
\Bigg].
\label{squarklattice}
\end{align}
\vspace{-0.15cm}
The coefficients $c_{-2}$, $c_{-1}$, $c_{m_\psi^2}$, and $c_{q^2}$ depend on the overlap parameters. The terms proportional to inverse powers of the lattice spacing give rise to a critical mass for the squark fields. For $r^{\psi}=\rho^{\psi}=r^{\lambda}=\rho^{\lambda}=1$, the critical mass is given by:
\be
m^{2\,\text{squark}}_{\text{crit.}} =
- \frac{g^2\,C_F}{16\,\pi^2}\begin{pmatrix}
1 & 0 \\
0 & 1
\end{pmatrix}
\left[30.8596 \,\frac{1}{a}\,m_\psi \, 
-77.2913  \frac{1}{a^2}\right].
\ee
Note that the critical squark mass is diagonal, and thus it introduces no mixing between the different squark components. The above value of the critical squark mass can be used as a ball-park estimate to calibrate Monte Carlo simulations and to appropriately tune the mass parameter in the lattice Lagrangian.
The squark field renormalization factor is determined by:
\be
\big\langle \tilde A(q)\, \tilde A^{\dagger}(q') \big\rangle^{\MSbar}_{\rm inv} = \left[Z_A^{LR,\MSbar}\right]^{-1/2}
\big\langle \tilde A(q)\, \tilde A^{\dagger}(q') \big\rangle^{LR}_{\rm inv} \left(\left[Z_A^{LR,\MSbar}\right]^{-1/2}\right)^\dagger,
\label{sqCond}
\ee
and the corresponding mass renormalization factor is defined by: ${{m_A^R}}^\dagger m_A^R =
\left(Z_{m_A}^{LR,\MSbar}\right)^\dagger {{m_A^B}}^\dagger m_A^B \,Z_{m_A}^{LR,\MSbar}$. Our results for the renormalization factors of the squark field and mass
in the $\overline{\mathrm{MS}}$ scheme, obtained at one-loop order in lattice
perturbation theory, are presented below for
$\rho^\lambda = \rho^\psi = r^\lambda = r^\psi = 1$:
\be
Z_A^{LR,\MSbar} = \openone + \frac{g^2\,C_F}{16\,\pi^2}\left[
\left(
-16.5576 + 3.79201\,\alpha +(1-\alpha)\log(a^2 \bar\mu^2)
\right)\begin{pmatrix}1 & 0 \\ 0 & 1\end{pmatrix}
+ \frac{2}{3}\begin{pmatrix}0 & 1 \\ 1 & 0\end{pmatrix}
\right],
\ee

\be
Z_{m_A}^{LR,\MSbar} = \openone - \frac{g^2\,C_F}{16\,\pi^2}\left[
\left(
-17.1764 - 2\,\log(a^2 \bar\mu^2)
\right)\begin{pmatrix}1 & 0 \\ 0 & 1\end{pmatrix}
+ \frac{2}{3}\begin{pmatrix}0 & 1 \\ 1 & 0\end{pmatrix}
\right].
\ee
Note that the off-diagonal contributions to $Z_{A}^{LR,\MSbar}$ and $Z_{m_A}^{LR,\MSbar}$ originate from the $\MSbar$-renormalized two-point Green's function appearing on the left-hand side of Eq.~\eqref{sqCond}. The renormalization factor $Z_{m_A}^{LR,\MSbar}$ is gauge independent, as is the quark-mass renormalization constant $Z_{m_\psi}^{LR,\MSbar}$, as expected.

\begin{figure}[ht!]
\centering
\includegraphics[scale=0.3]{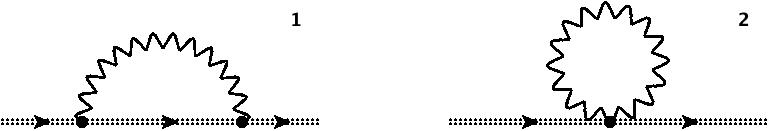}
\caption{One-loop Feynman diagrams contributing to the two-point Green's function $\langle c \, \bar c \rangle$. Notation is identical to that of Figure \ref{fig2ptGG}.}
\label{fig2ptcbarc}
\end{figure}

Lastly, the ghost field renormalization constant $Z_c$ (which will be used in the renormalization of coupling constants) is extracted from the one-loop ghost propagator. Summing the diagrams shown in Fig.~\ref{fig2ptcbarc}, we obtain:
\be
\big\langle c^{\alpha}(q),\bar{c}^{\beta}(q') \big\rangle^{LR}_{\rm inv} =
(2\pi)^4 \delta(q-q')\,\delta^{\alpha\beta}
\left[q^2-\Sigma_c^{\mathrm{latt}}(q)\right],
\ee
where: $\Sigma_c^{\mathrm{latt}}(q) =
\frac{g^2 N_c}{16\pi^2}\,q^2
\left[
4.6086-1.2029\,\alpha
-\frac{1}{4}(3-\alpha)\log\left(a^2q^2\right)
\right]$.
Imposing the $\MSbar$ renormalization condition yields:
\be
Z_c^{LR,\MSbar} =
1-\frac{g^2N_c}{16\pi^2}
\left[
3.6086-1.2029\,\alpha
-\frac{1}{4}(3-\alpha)\log\left(a^2\bar{\mu}^2\right)
\right].
\ee
\vspace{-1.5cm}
\section{Discussion and outlook}
\label{sec:conclusions}
We have constructed a lattice formulation of SQCD based on overlap fermions
that preserves a modified chiral symmetry. This formulation is expected to
require a reduced set of counterterms and allows for the computation of the
renormalization factors of all fields and masses in the theory. Future work will focus on determining the counterterms for the gauge, Yukawa,
and quartic scalar couplings. We will also investigate alternative squark
discretizations which may eliminate, or reduce, the required tuning for the critical squark mass.
\vspace{-0.5cm}
\section*{Acknowledgements}
\vspace{-0.5cm}
The project EXCELLENCE/0524/0208 is implemented under the programme of social cohesion ``THALIA 2021-2027'' co-funded by the European Union, through the Research and Innovation Foundation of Cyprus.

\end{document}